\documentclass[aps,pre,twocolumn,showpacs,superscriptaddress,groupedaddress]{revtex4-1} 

\usepackage{graphicx}        
\usepackage{dcolumn}        
\usepackage{bm}             
\usepackage{amssymb}
\usepackage{amsmath}        
\usepackage{color}
\definecolor{korr_26Apr}{rgb}{0,0,0} 
\definecolor{red}{rgb}{1,0,0}

\begin{document}

\widetext

\title{Numerical modeling of the wind flow over a transverse dune}
\author{Asc\^anio D. Ara\'ujo$^1$, Eric J. R. Parteli$^2$, Thorsten P\"oschel$^2$, Jos\'e S. Andrade Jr.$^1$ and Hans J. Herrmann$^{1,3}$}
\address{
~${\mbox{1.~Departamento de F\'{\i}sica, Universidade Federal do Cear\'a, Campus do Pici, 60451-970 Fortaleza, Cear\'a, Brazil.}}$ \\
~${\mbox{2.~Institute for Multiscale Simulation, Universit\"at Erlangen-N\"urnberg, N\"agelsbachstra{\ss}e~49b, 91052 Erlangen, Germany.}}$ \\
3.~~Computational Physics, IfB, ETH Z\"urich, Schafmattstra{\ss}e~6, 8093 Z\"urich, Switzerland.
}


\begin{abstract}
Transverse dunes, which form under unidirectional winds and have fixed profile in the direction perpendicular to the wind, occur on all celestial objects of our solar system where dunes have been detected. Here we perform a numerical study of the average turbulent wind flow over a transverse dune by means of computational fluid dynamics simulations. We find that the length of the zone of recirculating flow at the dune lee --- the {\em{separation bubble}} --- displays a surprisingly strong dependence on the wind shear velocity, $u_{\ast}$: it is nearly independent of $u_{\ast}$ for shear velocities within the range between $0.2\,$m$/$s and $0.8\,$m$/$s but increases linearly with $u_{\ast}$ for larger shear velocities. Our calculations show that transport in the direction opposite to dune migration within the separation bubble can be sustained if $u_{\ast}$ is larger than approximately $0.39\,$m$/$s, whereas a larger value of $u_{\ast}$ (about $0.49\,$m$/$s) is required to initiate this reverse transport.
\end{abstract}

\maketitle

\begin{figure*}[ht]
\begin{center}
\includegraphics[width=0.75 \textwidth]{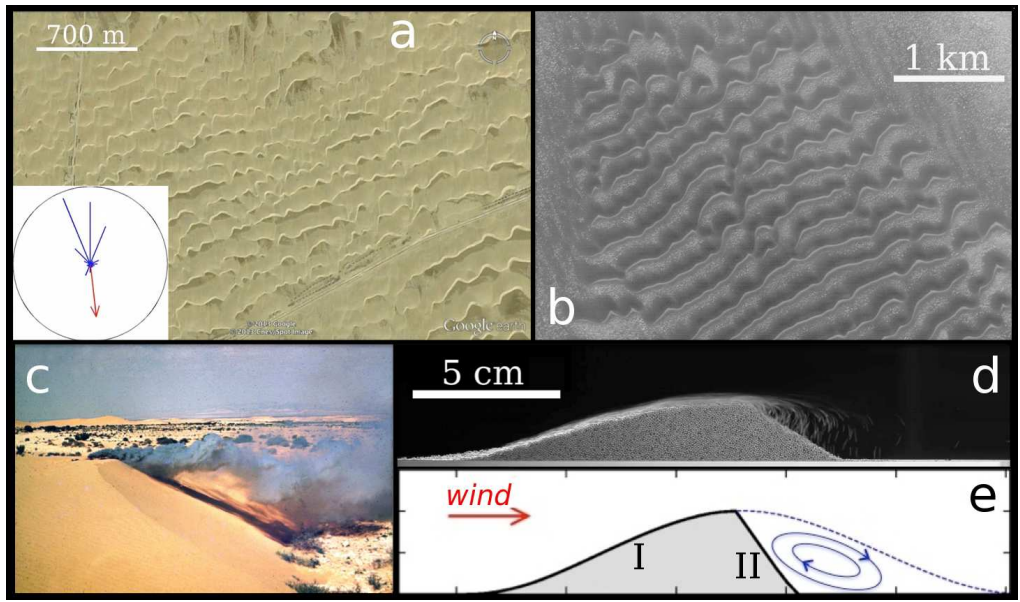}
\caption{{\bf{Flow separation at the lee of a transverse dune}}. {\bf{(a)}} Transverse dunes in Bahrein, near $25^{\circ}49$N, $49^{\circ}55$E (image credit: Google Earth). North is at top. In the bottom-left-hand-corner we see the sand rose for Doha AP in Qatar, located about $170\,$km SE of the transverse dune field. The vector length of each direction ($i$) of the sand rose gives the potential rate of sand transport due to sand-moving winds blowing from that direction, i.e. the drift potential, ${\mbox{DP}}_i = \sum U^2[U-U_{\mathrm{t}}]f_U$ \citep{Fryberger_and_Dean_1979,Tsoar_2001}. In this equation, $U$ is the wind velocity, $U_{\mathrm{t}}$ is the threshold wind speed for transport and $f_U$ is the fraction of time the wind speed exceeds $U_{\mathrm{t}}$. The arrow indicates the resultant drift potential, RDP $= \sum {\mbox{DP}}_i$, where the summation is over all directions of the sand rose. {\bf{(b)}} Transverse dunes on Mars, in a crater in Noachis Terra, near $41.7^{\circ}$S, $319.8^{\circ}$W (image courtesy of NASA/JPL/MSSS). {\bf{(c)}} Experiment with smoke over a barchan dune in Sinai for visualizing the flow pattern at the dune lee (image credit: Haim Tsoar). {\bf{(d)}} Longitudinal profile of a subaqueous transverse dune produced in a water tank (image courtesy of Ingo Rehberg; see Ref.~\cite{Groh_et_al_2008} for details about the experiments). The direction of the water stream is from left to right. The recirculating pattern of the flow at the dune lee is visible. {\bf{(e)}} Sketch of the longitudinal profile of a transverse dune indicating its windward side (I), its slip face (II), the separation streamline (dashed line) connecting the dune brink with the ground, as well as the recirculating flow pattern within the ``separation bubble'' at the dune lee.}
\label{fig:images}
\end{center}
\end{figure*}

The shape of dunes depends fundamentally on flow directionality and availability of mobile sediment. In areas of constant wind direction, crescent-shaped {\em{barchans}} are the characteristic dune type, provided the amount of sand is sufficiently low such that the dunes occur on top of bedrock and well separated from each other \citep{Bagnold_1941,Bourke_and_Goudie_2009}. Under higher sand availability, the {\em{transverse dune}} (Fig.~\ref{fig:images}a,b) --- which has nearly fixed profile in the direction perpendicular to the wind --- is the prevailing dune type \citep{Frank_and_Kocurek_1996}. Moreover, longitudinal {\em{seif dunes}}, which display a characteristic meandering and elongate into the resultant transport direction, form under bimodal wind regimes, while accumulating {\em{star dunes}} occur in areas where the wind blows from multiple directions \citep{Pye_and_Tsoar_1990}.

Numerical modeling can offer a helpful tool for dune research as long time-scale processes associated with the dynamics of dunes and dune fields can be efficiently simulated on the computer \citep{Sauermann_et_al_2001,Kroy_et_al_2002}. Indeed, in order to model dune morphodynamics, an analytical description of the average turbulent wind flow over the sand terrain is required. Such a model has been developed in Ref.~\cite{Jackson_and_Hunt_1975} for computing the shear stress over a topography consisting of smooth hills, and has been further improved by many authors, in particular by Weng et al.~\citep{Weng_et_al_1991}. However, this analytical model is not applicable if the dune has a sharp brink. At the brink of the dune, flow separation occurs, thereby leading to a zone of recirculating flow at the dune lee (cf.~Fig.~\ref{fig:images}c,d) which is not included in the analytical model. In order to overcome this problem, dune simulations are performed by accounting for an {\em{ad hoc}} {\em{separating streamline}} enclosing the area of recirculating flow (or ``separation bubble'') at the dune lee and connecting the brink to the ground at the reattachment point downwind. First, the average shear stress field above the smooth envelope comprising the separating streamline and the windward side of the dune is computed. Thereafter, the shear stress within the separation bubble is set to zero, since net transport within the zone of recirculating flow essentially vanishes. However, the main factors controlling the shape of the separating streamline, which affects the shear stress on the windward side, are still uncertain.

Computational Fluid Dynamics simulations can be employed in order to solve the average Navier-Stokes equations for the turbulent wind field over dunes \citep{Wiggs_2001}. One typically focuses on the flow over transverse dunes, as this dune type is the simplest one: since its profile is nearly invariant in the direction orthogonal to the wind, simulations can be performed by considering only the along-wind (longitudinal) profile of the dune \citep{Walker_and_Nickling_2002,Walker_and_Nickling_2003,Parsons_et_al_2004}. Moreover, the longitudinal profile of the transverse dune provides a good approximation for the central slice of a crescent-shaped barchan of similar height, since lateral flux along the symmetry axis of a barchan is neglibible \citep{Duran_et_al_2010}. 

Previous modeling focused on the relation between dune shape or interdune spacing and the characteristics of the flow over the dune profile. Different functional forms have been proposed to describe the dependence of the separation streamline, $s(x)$, on the downwind position ($x$), including third-order polynomials \citep{Sauermann_et_al_2001,Kroy_et_al_2002,Paarlberg_et_al_2007} and ellipses \citep{Schatz_and_Herrmann_2006}. The parameterization is determined by fitting the curve of $s(x)$ to simulations considering different dune sizes and also dune crests of different shapes (different crest-brink distances). Note that the shape of the streamline can be different for isolated or closely spaced dunes \citep{Schatz_and_Herrmann_2006,Paarlberg_et_al_2007}. However, the effect of the wind speed on the separation streamline is still uncertain. On the one hand, information about the wind shear velocity, $u_{\ast}$, is encoded in the shape of the transverse dune. Modeling has shown that the lower $u_{\ast}$, the rounder the dune crest \citep{Kroy_et_al_2005}, whereas it is known that the zone of recirculating flow becomes smaller the rounder the dune crest \citep{Herrmann_et_al_2005,Schatz_et_al_2006}. On the other hand, the fit parameters for determining $s(x)$ for different dune profiles are typically obtained from simulations performed with one single value of wind shear velocity regardless of the dune shape.

In the present work, we use Computational Fluid Dynamics simulations in order to calculate the average turbulent wind field over a transverse dune of fixed profile under different flow speeds. Specifically, we consider an isolated transverse dune of fixed profile and investigate how the shape and the size of the separation bubble, as well as the characteristics of the flow inside it, vary with the average wind shear velocity, $u_{\ast}$, to which the dune is subjected. 

\begin{figure}[ht]
\begin{center}
\includegraphics[width=1.0 \columnwidth]{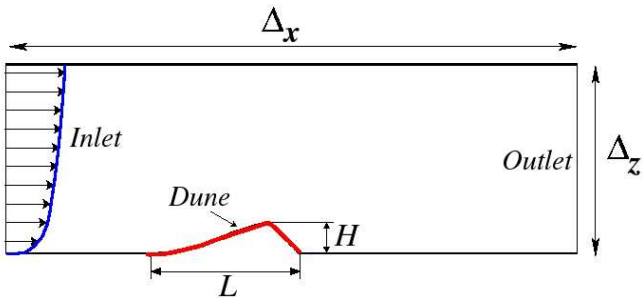}
\caption{{\bf{Schematic diagram of the computational setup}}. The simulation box has width ${\Delta}_x=100\,$m and height ${\Delta}_z=20\,$m, while the dune has height $H=3.26 \text{m}$ and total along-wind width $L=25 \text{ m}$. The wind blows from left to right and the wind velocity increases with the height following Eq.~(\ref{eq:wind_profile}).} 
\label{fig:simulation_box}
\end{center}
\end{figure}

Figure \ref{fig:simulation_box} shows the schematic representation of the setup employed in our calculations. The dune, placed on top of the bottom wall of a two-dimensional channel of height $\Delta_z = 20\,$m and width $\Delta_x = 100\,$m, has height $H = 3.26\,$m at the brink and width $L = 25\,$m in the direction of the wind ($x$). The dune is placed in the center of the box, and the dimensions of the box are chosen in such a way that the dune is far enough from the channel's walls, thus avoiding that the results are affected by border effects. That is, increasing the size of the box does not change the results of our calculations. The wind velocity $u(z)$ increases logarithmically with the height $z$ above the bed level ($h$) \citep{Bagnold_1941,Pye_and_Tsoar_1990}. That is,
\begin{equation}
u(z) = \frac{u_{\ast}}{\kappa}{\mathrm{log}}{\frac{z-h}{z_0}}, \label{eq:wind_profile}
\end{equation}
where $\kappa = 0.4$ is the von K\'arm\'an constant, $z_0$ is the surface roughness and $u_{\ast}$ is the upwind shear velocity of the wind, which is used to define the (upwind) shear stress, 
\begin{equation}
\tau = {\rho}_{\mathrm{f}}{u_{\ast}^2}, \label{eq:tau}
\end{equation}
with ${\rho}_{\mathrm{f}} = 1.225\,$kg$/$m$^3$ standing for air density. In the present study, we use a value of $z_0 = 100\,{\mu}$m. This value has been obtained in a previous work \citep{Almeida_et_al_2006} by fitting Eq.~(\ref{eq:wind_profile}) to the steady-state wind profile within the simulation box (Fig.~\ref{fig:simulation_box}), which has been generated  upon imposing a pressure difference between the extremities of the ``wind tunnel'' using different values of induced flow speed \citep{Almeida_et_al_2006}. We find that using other values of $z_0$ within the range between $10\,{\mu}$m and $1.0\,$mm \citep{Pye_and_Tsoar_1990,Kok_et_al_2012} has a negligible effect on the values of shear stress obtained in our calculations.

While previous works investigated the flow over triangular obstacles \citep{Parsons_et_al_2004} or artificial dune profiles with a windward side built using half-circles \citep{Schatz_and_Herrmann_2006}, the dune profile used in our calculations is generated using a morphodynamic model for dune formation \citep{Sauermann_et_al_2001,Kroy_et_al_2002,Sauermann_et_al_2003,Parteli_et_al_2006,Duran_et_al_2010}. Using this model, we obtain the longitudinal profile of a transverse dune by starting with a Gaussian hill subjected to a wind shear velocity $u_{\ast} = 0.4\,$m$/$s, which is a typical value for the average shear velocity of sand-moving winds on Earth's dune fields \citep{Fryberger_and_Dean_1979}. The final dune shape (shown in Fig.~\ref{fig:simulation_box}) has a sharp brink which separates the windward side, on which wind-driven particle transport --- that is, saltation \citep{Bagnold_1941,Makse_2000} --- occurs, from the slip-face, which has the inclination of the angle of repose $34^{\circ}$ and is the side of the dune where avalanches occur (see Fig.~\ref{fig:images}e). 

Motivated by the observation that some dunes on Mars might evolve from sand deposition in the wake of an indurated dune \citep{Schatz_et_al_2006}, here we aim to understand how the size of the separation bubble of a transverse dune of fixed profile changes when the wind speed is varied. Therefore, in the present study the same dune profile of Fig.~\ref{fig:simulation_box} is used to perform all simulations, regardless of the value of the wind velocity.

As mentioned previously, one further aim of the present study is to understand how the flow characteristics inside the separation bubble of a transverse dune changes upon a variation in wind speed. We note that the steady-state profile of a mobile transverse dune is adapting to the flow conditions within the time-scale of dune motion. A systematic investigation of the role of the average wind speed for the long-term evolution of the dune shape would thus require computational fluid dynamic calculations coupled with a morphodynamic model which accounts for the changes in the dune profile due to the wind-driven saltation flux. On the other hand, the time-scale associated with the reconstitution of the dune shape after a change in flow conditions --- that is, the dune turnover time, $T_{\mathrm{m}}$ --- is of the order of months or years \citep{Pye_and_Tsoar_1990,Parteli_et_al_2011,Kok_et_al_2012}, and is thus much larger than the time-scale of the turbulent fluctuations of the wind flow over the dune profile (of the order of seconds \citep{Duran_et_al_2011}). Thus, keeping the dune profile fixed in the simulations provides an approximation for investigating how changes in the upwind shear velocity occurring at time-scales much smaller than the dune turnover time affect the nature of the flow inside the dune separation bubble \citep{Walker_and_Nickling_2002,Walker_and_Shugar_2013}.

\section*{Results\label{sec:results}}

\subsection*{\label{sec:sepbub_size}Length of the separating streamline}

We calculate the size of the flow separation zone for different values of $u_{\ast}$ within the range $0.01\,$m$/$s $< u_{\ast} < 1.6\,$m$/$s.

It is useful to express this range of shear velocities in terms of the Reynolds number (Re). In order to define Re, we take the dune height as the characteristic length-scale associated with the flow over the transverse dune \citep{Sauermann_2001}. Therefore, we write the Reynolds number as ${\mbox{Re}} = U_1{H}{\rho}_{\mathrm{f}}/{\eta}$, where $H \approx 3.26\,$m for the transverse dune investigated here, while $\eta \approx 1.78 \times 10^{-5}\,$kg\,m$^{-1}$s$^{-1}$ is the dynamic viscosity of the air, and $U_1$ is a characteristic value of the wind velocity --- which is typically taken at a height of 1 m above the bed \citep{Fryberger_and_Dean_1979}. Using Eq.~(\ref{eq:wind_profile}) to calculate $U_1 = u(z=1\,$m$)$, we obtain that the above range of $u_{\ast}$ corresponds to Reynolds numbers in the interval $5.2 \times 10^4 < {\mbox{Re}} <  8.3 \times 10^6$. Therefore, all values of $u_{\ast}$ used in our simulations are associated with fully developed turbulent wind flows, since Reynolds numbers larger than about $6000$ lead to turbulent flows in the atmospheric boundary layer \citep{Houghton_1986}.

\begin{figure}[htpb]
\begin{center}
\includegraphics[width=1.0 \columnwidth]{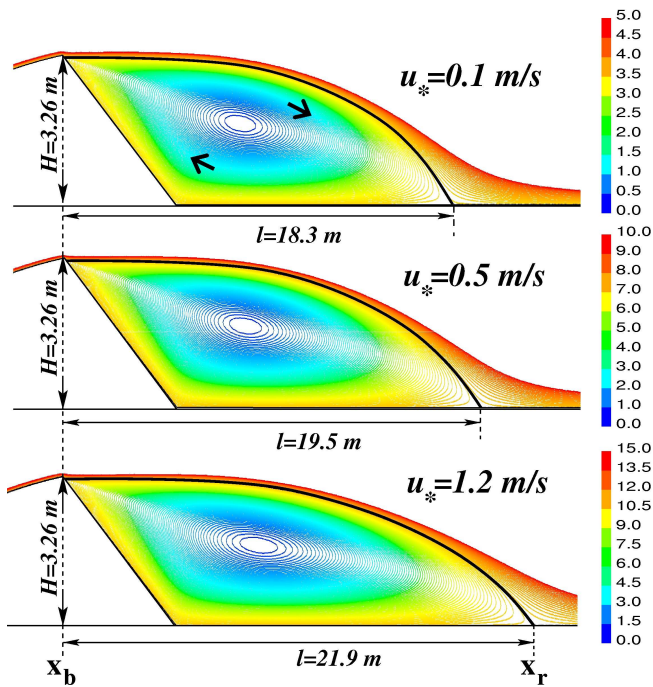}
\caption{{\bf{Calculation of the average turbulent wind flow over the longitudinal slice of the transverse dune}}. The calculation is performed for different values of wind shear velocity $u_{\ast}$. From top to bottom, the values of $u_{\ast}$ used in the calculations are $0.1\,$m$/$s, $0.5\,$m$/$s and $1.2\,$m$/$s. The dune has height $H = 3.26\,$m at the brink. Colors indicate mass flux per unit time ($q$), associated with each streamline, in units of $\text{kg/s}$, whereas the value of $q$ increases from blue to red. The separating streamline is denoted by the continuous, thick line in each figure. The point where the separation streamline touches the ground is the flow reattachment point ($x_r$), while the horizontal distance from the dune's brink (which is at position $x_b$) to $x_r$ gives the reattachment length ($l$) associated with the given value of $u_{\ast}$. The mass flow along the recirculating streamlines inside the separation bubble is in the clockwise direction, as indicated by the black arrows in the plot corresponding to $u_{\ast} = 0.1\,$m$/$s (top). The mass flow along all streamlines above the separating streamline is in the forward direction.}
\label{fig:velocity_field}
\end{center}
\end{figure}

Figure \ref{fig:velocity_field} shows the contour plot of the mass flux in the wake of the dune for different values of $u_{\ast}$. For all values of $u_{\ast}$ within the aforementioned range, a well-developed zone of recirculating flow is observed downwind of the dune. This zone, also called ``separation bubble'', separates recirculation from the upper region of laminar flow. The streamline which separates both zones, i.e. the separating streamline, connects the brink to the ground at the reattachment point. The reattachment point is located at position $x_{\mathrm{r}}$, which has a horizontal distance $l$ from the brink position ($x_{\mathrm{b}}$). We investigate how the size of the separation zone and the shape of separating streamline change with $u_{\ast}$. 

\begin{figure*}[!htpb]
\begin{center}
\includegraphics[width=1.0 \textwidth]{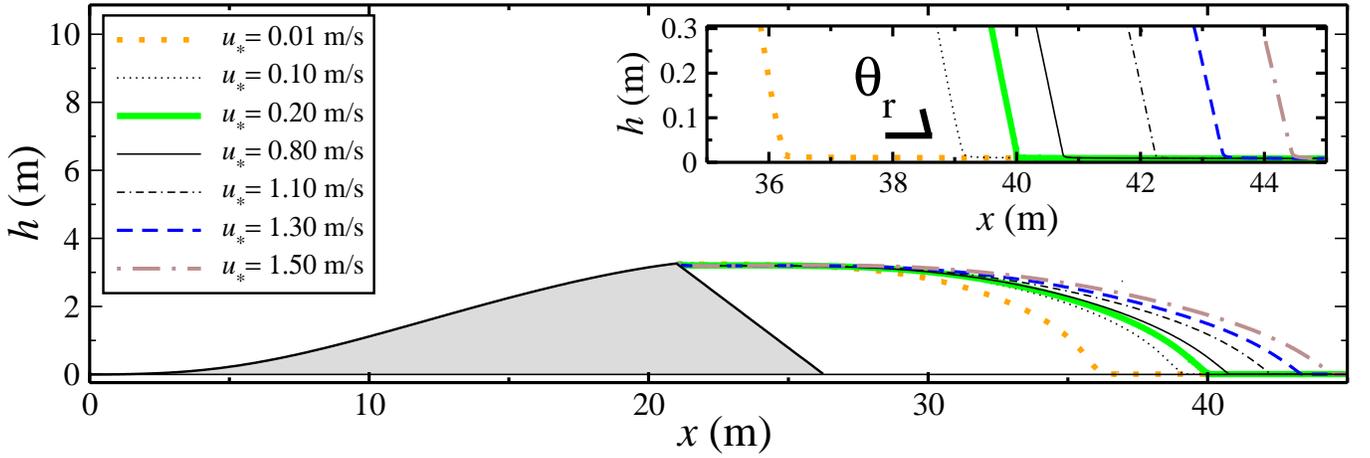}
\caption{{\bf{The length of the separation streamline increases with the upwind shear velocity.}} Main plot: Schematic diagram showing the separation streamlines of the transverse dune calculated using different values of shear velocity $u_{\ast}$. Wind blows from left to right, and the longitudinal slice of the transverse dune is represented by the grey area. The continuous streamlines in the intermediate range of $u_{\ast}$ displayed in the figure denote the lower and upper bonds ($u_{\ast} = 0.2$ m$/$s and $0.8$ m$/$s, respectively) of the range of $u_{\ast}$ within which the reattachment length $l$ is approximately constant and equal to six times the dune height (cf.~Fig.~\ref{fig:l_versus_ustar}). In the inset we see that the separation streamlines make an angle ${\theta}_{\mathrm{r}}$ with the horizontal at the reattachment point, which depends on $u_{\ast}$ as depicted in the inset of Fig.~\ref{fig:l_versus_ustar}.}
\label{fig:streamlines}
\end{center}
\end{figure*}

The position $x_{\mathrm{r}}$ of the reattachment point is defined as the downwind position at which the horizontal component of the wind velocity at the ground changes sign from upwind ($u < 0$) to downwind ($u > 0$). In other words, the separation streamline is the streamline associated with the smallest mass flux (where the mass flux is in units of $\text{kg/s}$ and increases from blue to red in Fig.~\ref{fig:velocity_field}) which does not form a loop. In Fig.~\ref{fig:velocity_field}, the streamline associated with each value of $u_{\ast}$ is represented by the continuous black line connecting the brink to the ground at position $x_{\mathrm{r}}$. We see that the length of flow separation, ${l} = x_{\mathrm{r}}-x_{\mathrm{b}}$, increases with $u_{\ast}$. This is shown more clearly in the main plot of Fig.~\ref{fig:streamlines}, which displays the separating streamlines corresponding to different values of $u_{\ast}$. We also see in the inset of this figure that the streamlines make an angle ${\theta}_{\mathrm{r}}$ with the horizontal at the reattachment point, which is different from zero. The fact that the separating streamline makes an angle with the bed at the reattachment point was also observed in previous numerical studies \citep{Schatz_and_Herrmann_2006}, as well as in experiments on subaqueous dunes \citep{Paarlberg_et_al_2007}. 

\begin{figure}[!htpb]
\begin{center}
\includegraphics[width=1.0 \columnwidth]{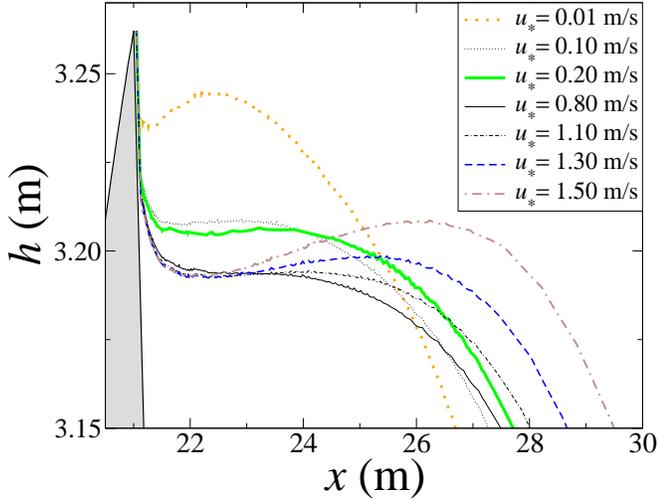}
\caption{{\bf{Flow separation does not occur at the brink position.}} The figures shows the close-up of the separation streamlines displayed in Fig.~\ref{fig:streamlines}, now displaying their behavior near the brink of the dune (grey area).}
\label{fig:zoom}
\end{center}
\end{figure}

At the brink of the dune, the separating streamline displays a non-trivial behaviour as shown in Fig.~\ref{fig:zoom}. As we can see in this figure, the streamline does not separate from the dune profile at the brink, but at a small distance down the dune slip face. We find that the streamline is about two grid spacings apart the brink position, independently of the grid spacing used. The origin of this behavior is that the steep gradients of the fluid flow at the brink make the simulation innacurate at the brink, and since the grid spacing is a variable associated with the simulation rather than the Navier-Stokes equations describing the flow, we conclude that the dip in the separating streamline near the brink shown in Fig.~\ref{fig:zoom} is a numerical artifact. In a previous work \citep{Schatz_and_Herrmann_2006}, it was shown that, by considering the part of the separating streamline which curves downwards and neglecting the dip near the brink, the separating streamline can be well fitted by ellipses connecting the brink to the reattachment point at the ground.

\begin{figure}[!htpb]
\begin{center}
\includegraphics[width=1.0 \columnwidth]{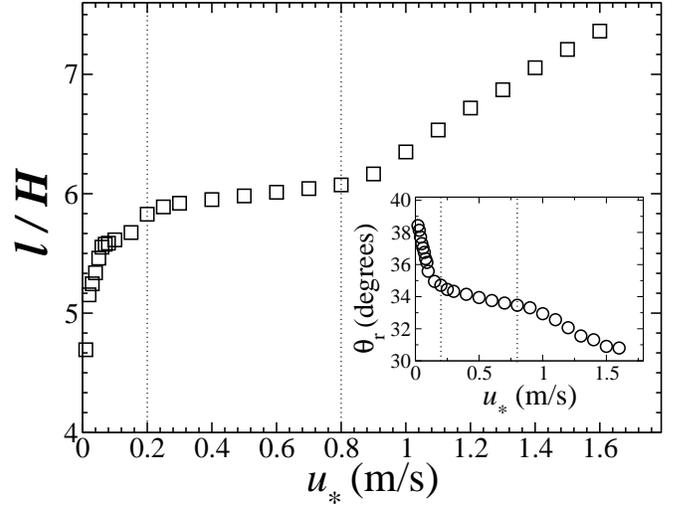}
\caption{{\bf{Effect of the wind shear velocity on the size and the shape of the separation streamline.}} The main plot shows the dimensionless reattachment length $l/H$ (where $H \approx 3.26$~m is the height of the dune) as a function of the wind shear velocity $u_{\ast}$. The region between the vertical dashed lines denotes the range of $u_{\ast}$ within which $l \approx 6\,H$. The inset shows the angle ${\theta}_{\mathrm{r}}$ (defined in Fig.~\ref{fig:streamlines}) which the streamlines associated with different values of $u_{\ast}$ make with the horizontal at the reattachment point.}
\label{fig:l_versus_ustar}
\end{center}
\end{figure}

In the main plot of Fig.~\ref{fig:l_versus_ustar} we see the dimensionless reattachment length $l/H$, where $H$ is the dune height at the brink, for different values of $u_{\ast}$. We identify three different regimes for the growth of $l/H$ with $u_{\ast}$. The first regime (I) corresponds to $u_{\ast}$ below $0.2\,$m$/$s. The second regime (II) is associated with an intermediate range of $u_{\ast}$ between $0.2\,$m$/$s and $0.8\,$m$/$s, while values of $u_{\ast}$ above this range characterize regime III. We see that for values of $u_{\ast}$ within regime II (intermediate $u_{\ast}$), the increase of $l$ with $u_{\ast}$ is small. In this regime, $l$ is around six times $H$, which is roughly the value observed in previous computational fluid dynamics simulations of the wind flow over isolated transverse dunes \citep{Parsons_et_al_2004,Schatz_and_Herrmann_2006}. However, in both regimes I (low $u_{\ast}$) and III (high $u_{\ast}$) the reattachment length increases with $u_{\ast}$ much faster than it does within regime II.

It is interesting to note that regime II includes the range of $u_{\ast}$ that is typical for average sand-moving winds on terrestrial dune fields. In order for aeolian sand transport to occur, the average wind speed must be larger than the threshold for sustained sand transport, $u_{\mathrm{t}}$, which for terrestrial dunes is about $0.2\,$m$/$s \citep{Bagnold_1941} and thus roughly matches the lower limit of the $u_{\ast}$ interval corresponding to regime II. Moreover, the upper limit of regime II approximately coincides with the threshold for transport of the sand particles through suspended load, which for terrestrial sand transport is about $4u_{\mathrm{t}}$ (see e.g.~Ref.~\cite{Kok_et_al_2012}). Indeed, since average shear velocities associated with saltation transport and dune formation in terrestrial dune fields are typically smaller than $\approx 0.5\,$m$/$s \citep{Fryberger_and_Dean_1979,Parteli_et_al_2006,Kok_et_al_2012}, previous studies of flow separation were performed with $u_{\ast}$ values not exceeding this ``upper bound'' \citep{Walker_and_Nickling_2003,Schatz_and_Herrmann_2006}. 

The dependence of the streamline's slope ${\theta}_{\mathrm{r}}$ at the reattachment point on $u_{\ast}$ is shown in the inset of Fig.~\ref{fig:l_versus_ustar}. We see that ${\theta}_{\mathrm{r}}$ is a decreasing function of $u_{\ast}$ which means that the angle at the reattachment point is larger the smaller the length of the separation zone. Our simulations indicate that, for values of $u_{\ast}$ corresponding to the average shear velocities of Earth's sand-moving winds (between $0.2\,$m$/$s and $0.5\,$m$/$s), ${\theta}_{\mathrm{r}}$ is about $34^{\circ}$, which, interestingly, approximately coincides with the angle of repose of sand (the slope of the slip-face). 

\subsection*{Flow characteristics within the separation bubble}

In addition to characterizing the separation streamline, it is interesting to study the role of the effect of the wind speed on the characteristics of the flow within the zone of recirculating flow. 

\begin{figure}[htpb]
\begin{center}
\includegraphics[width=1.0 \columnwidth]{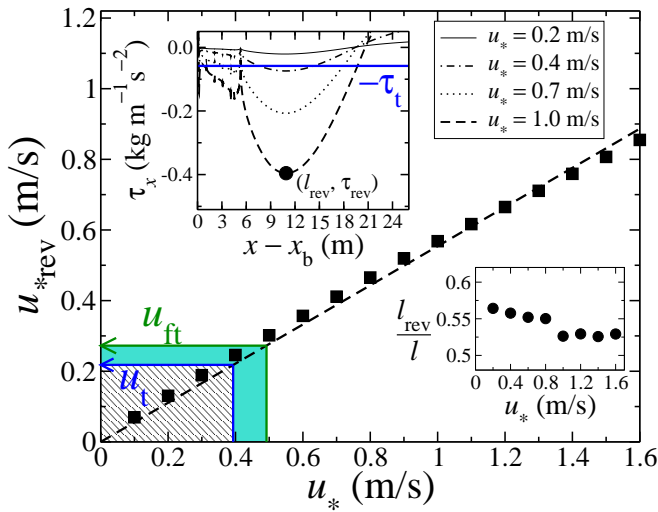}
\caption{{\bf{The shear velocity associated with the reversing flow at the ground within the separation bubble scales linearly with the upwind shear velocity.}} In the {\em{upper inset}}, we see the value of ${\tau}_x$ (cf.~Eq.~(\ref{eq:tau_x})) associated with the reversing flow within the separation bubble as a function of the downwind distance $x-x_{\mathrm{b}}$, where $x_{\mathrm{b}}$ is the brink position. The minimum of ${\tau}_x$, which gives the maximal magnitude of the shear stress associated with the reversing flow, is denoted by ${\tau}_{\mathrm{rev}}$ and occurs at a horizontal distance ${l}_{\mathrm{rev}}$ downwind of the brink. The blue horizontal line indicates the impact threshold shear stress, ${\tau}_{\mathrm{t}} = {\rho}_{\mathrm{fluid}}u_{\mathrm{t}}^2$. The black dot identifies the point $\{l_{\mathrm{rev}},{\tau}_{\mathrm{rev}}\}$ associated with $u_{\ast} = 1.0\,$m$/$s. We see in the {\em{lower inset}} that $l_{\mathrm{rev}} \approx 55\%$ of the reattachment length $l$ and is nearly independent of $u_{\ast}$. {\em{Main plot}}: Magnitude of the maximal shear velocity at the ground, $u_{{\ast}{\mathrm{rev}}} \equiv {\sqrt{|{\tau}_{\mathrm{rev}}|/{\rho}_{\mathrm{fluid}}}}$, as a function of the upwind shear velocity, $u_{\ast}$. The best fit using the equation $u_{{\ast}{\mathrm{rev}}} = ku_{\ast}$ (dashed line) gives $k \approx 0.55$. The dashed area indicates the range of $u_{\ast}$ for which no sustained transport occurs ($u_{{\ast}{\mathrm{rev}}} < u_{\mathrm{t}}$), while the regime of $u_{\ast}$ where $u_{{\ast}{\mathrm{rev}}}$ is smaller than the threshold for direct entrainment ($u_{{\mathrm{ft}}}$) is denoted by the green filled area.}
\label{fig:ustar_sepbub}
\end{center}
\end{figure}

One aspect which is missing in practically all models of dune formation is that, at the lee of the dune, sand can be transported in the direction {\em{opposite}} to the wind due to the recirculating flow in the separation bubble. As a matter of fact, since net transport within the wake of the dune nearly vanishes, the shear stress within the separation bubble in dune models is usually set as zero \citep{Momiji_et_al_2000,Kroy_et_al_2002,Schwaemmle_and_Herrmann_2004,Pelletier_2009,Zheng_et_al_2009,Duran_et_al_2010}. However, the shape of dunes at the lee side can be affected by sand transport due to the recirculating wind patterns \citep{Pye_and_Tsoar_1990,Tsoar_2001}. How does the strength of the reverse flow depend on the upwind wind speed? In order to address this question, we calculate the bed shear stress at the ground within the zone of recirculating flow and 
in particular we study how the magnitude of the shear velocity associated with the reversing flow within the separation bubble depends on the upwind wind shear velocity, $u_{\ast}$. The results of our calculations are shown in Fig.~\ref{fig:ustar_sepbub} and discussed in the paragraphs which follow.

By neglecting the lateral component of the shear stress (that is, the one parallel to the axis of the transverse dune), we can write,
\begin{equation}
{\vec{\tau}} = {\tau}_x{\hat{\mbox{e}}}_x, \label{eq:tau_x}
\end{equation}
where ${\hat{\mbox{e}}}_x$ is the unit vector that points in wind direction. In the upper inset of Fig.~\ref{fig:ustar_sepbub} we show the value of ${\tau}_x$ within the separation bubble as a function of the horizontal distance $x$ relative to the brink position $x_{\mathrm{b}}$ for different upwind shear velocities $u_{\ast}$. In the simulations, ${\tau}_x$ is computed with Eq.~(\ref{eq:tau}) by using the local wind shear velocity $u_{\ast}$ associated with the velocity profile $u_x(z)$ (where $u_x$ is the component of the wind velocity in the $x$ direction). We see that ${\tau}_x$ has negative values within the separation bubble. The behavior of ${\tau}_x$ near the brink is affected by the strong gradients of the flow resulting from the discontinuity of the dune surface at the brink, as discussed previously. Moroever, the behavior of ${\tau}_x$ inside the separation bubble is sensitive to the strong gradient of $h(x)$ that occurs at the position of the slip face foot, that is, at $x = x_{\mathrm{f}} \approx 26.3\,$m. As we can see in the upper inset of Fig.~\ref{fig:ustar_sepbub}, for all values of $u_{\ast}$, ${\tau}_x$ has a dip at that position (that is, at $x_{\mathrm{f}} - x_{\mathrm{b}} \approx 5.3\,$m). After position $x_{\mathrm{f}}$, ${\tau}_x$ curves downward until it achieves a minimal value.  We define $l_{\mathrm{rev}}$ as the distance $x-x_{\mathrm{b}}$ associated with the smallest value of ${\tau}_x$ inside the separation bubble, which we call ${\tau}_{\mathrm{rev}}$. For instance, in the upper inset of Fig.~\ref{fig:ustar_sepbub}, the point ($l_{\mathrm{rev}},{\tau}_{\mathrm{rev}}$) that corresponds to the upwind shear velocity $u_{\ast} = 1.0\,$m$/$s is denoted by a black dot. 

Note that while the dip of the separation streamline two grid spacings after the dune brink (cf.~Fig.~\ref{fig:zoom}) is a numerical artifact which occurs due to the flow separation at the sharp brink but is independent of the dune shape \citep{Schatz_and_Herrmann_2006}, the profile of the shear stress {\em{is}} affected by the local slope of the surface. That is, both the dip of the shear stress occurring at the downwind foot of the dune and the saddle of ${\tau}_x$ at position $x_{\mathrm{b}} + l_{\mathrm{rev}}$ (cf.~upper inset of Fig.~\ref{fig:ustar_sepbub}) are no numerical artifacts but result from the local gradients in the slope of the envelope $h(x)$ comprising the dune surface and the ground.

From the magnitude of ${\tau}_{\mathrm{rev}}$, we obtain the value of the shear velocity $u_{{\ast}{\mathrm{rev}}} = {\sqrt{|{\tau}_{\mathrm{rev}}|/{\rho}_{\mathrm{air}}}}$ that is associated with the reversing wind flow at the position $l_{\mathrm{rev}}$. The values of $u_{{\ast}{\mathrm{rev}}}$ obtained for different values of the upwind shear velocity $u_{\ast}$ are displayed in the main plot of Fig.~\ref{fig:ustar_sepbub} (squares). The dashed line denotes the best fit to the simulation data using the equation,
\begin{equation}
u_{{\ast}{\mathrm{rev}}} = ku_{\ast},
\end{equation}
which gives $k \approx 0.55$. The lower inset of Fig.~\ref{fig:ustar_sepbub} shows the ratio $l_{\mathrm{rev}}/l$ as a function of $u_{\ast}$. We see that the maximal value of $u_{{\ast}{\mathrm{rev}}}$ associated with the recirculating flow occurs at a downwind distance $l_{\mathrm{rev}}$ that is approximately equal to $55\%$ of the reattachment length $l$, whereas only a weak dependence of $l_{\mathrm{rev}}/l$ on $u_{\ast}$ is observed in our simulations.

Based on the results for $u_{{\ast}{\mathrm{rev}}}$ found from our simulations, we can estimate for which values of the upwind shear velocity $u_{\ast}$ surface transport of sand within the separation bubble can occur in the direction opposite to the dune migration trend. In order for sand transport to {\em{begin}}, the local wind shear velocity must exceed the so-called fluid threshold $u_{{\mathrm{ft}}}$. This threshold shear velocity is given by the equation \citep{Bagnold_1941,Kok_et_al_2012},
\begin{equation}
u_{{\mathrm{ft}}} = A{\sqrt{\frac{gd({\rho}_{\mathrm{p}}-{\rho}_{\mathrm{f}})}{{\rho}_{\mathrm{f}}}}}, \label{eq:u_ft}
\end{equation}
where $g$ is gravity, $d$ and ${\rho}_{\mathrm{p}}$ are the mean diameter and density of the sand grains, respectively, while $A$ is the Bagnold-Shields parameter, which depends on the shape and sorting of the grains and on the angle of internal friction \citep{Shields_1936,Sauermann_2001}. Furthermore, the value of $A$ depends on the attributes of sediment and fluid and can be estimated by numerically solving the following empirical equation derived by Iversen and White \cite{Iversen_and_White_1982},
\begin{equation}
A = 0.129\,\sqrt{{\frac{1+6.0 \times 10^{-7}/[{{\rho}_{\mathrm{p}}gd^{2.5}}]}{1.928{\mbox{Re}}_{{\ast}{\mathrm{ft}}}^{0.092}-1}}},
\end{equation}
for $0.03 \leq {\mbox{Re}}_{{\ast}{\mathrm{ft}}} < 10$, and,
\begin{eqnarray}
A &=& 0.129\,\sqrt{1+6.0 \times 10^{-7}/[{{\rho}_{\mathrm{p}}gd^{2.5}}]} \times \nonumber \\ && \left\{{1 - 0.0858\,{\mbox{exp}}[-0.0617({\mbox{Re}}_{{\ast}{\mathrm{ft}}}-10)]}\right\},
\end{eqnarray}
for ${\mbox{Re}}_{{\ast}{\mathrm{ft}}} \geq 10$, where ${\mbox{Re}}_{{\ast}{\mathrm{ft}}} = u_{\mathrm{ft}}d/{\nu}$ is the friction Reynolds number associated with the threshold shear velocity $u_{{\mathrm{ft}}}$, while the constant $6.0 \times 10^{-7}$ has units of kg$\,$m$^{0.5}$s$^{-2}$. By considering that the average grain size of the sand that composes terrestrial dunes is $d \approx 250\,{\mu}$m \citep{Pye_and_Tsoar_1990} and taking ${\rho}_{\mathrm{p}} \approx 2650\,$kg$/$m$^3$ (density of quartz), we obtain $A \approx 0.118$ and, 
\begin{equation}
u_{\mathrm{ft}} \approx 0.272\,{\mbox{m}}/{\mbox{s}}, \label{eq:u_ft_value}
\end{equation} 
which gives the minimal value of $u_{{\ast}{\mathrm{rev}}}$ required to {\em{initiate}} transport in the direction opposite to dune migration within the separation bubble. 

Indeed, once sand transport begins, it can be sustained at shear velocities lower than $u_{\mathrm{ft}}$ due to the splash mechanism resulting from the grain-bed collisions. The minimal shear velocity required to {\em{sustain}} saltation transport is the so-called impact threshold shear velocity, $u_{\mathrm{t}}$. As found experimentally and in numerical simulations, $u_{\mathrm{t}}$ is approximately equal to $80\%$ of $u_{\mathrm{ft}}$ \citep{Bagnold_1941,Kok_et_al_2012}. Therefore, using the values of $d$ and ${\rho}_{\mathrm{p}}$ associated with the characteristics of terrestrial dunes' sand specified above, we obtain,
\begin{equation}
u_{\mathrm{t}} \approx 0.218\,{\mbox{m}}/{\mbox{s}}. \label{eq:u_t_value}
\end{equation}
This value of shear velocity gives, thus, the minimal value of $u_{{\ast}{\mathrm{rev}}}$ required for reverse transport within the separation bubble to be sustained.   

We find that the values of upwind shear velocity $u_{\ast}$ associated with the minimal values of $u_{{\ast}{\mathrm{rev}}}$ required to initiate and sustain transport in the direction opposite to dune migration within the separation bubble are $u_{{\ast}{\mathrm{u}}} \approx 0.49\,$m$/$s and $u_{{\ast}{\mathrm{b}}} \approx 0.39\,$m$/$s, respectively. The dashed area in the main plot of Fig.~\ref{fig:ustar_sepbub} denotes the range of $u_{\ast}$ where no transport can occur ($u_{{\ast}{\mathrm{rev}}} < u_{\mathrm{t}}$). Furthermore, the green filled area identifies the regime of $u_{\ast}$ where transport within the bubble can be sustained, once initiated ($u_{\mathrm{t}} < u_{{\ast}{\mathrm{rev}}} < u_{{\mathrm{ft}}}$), whereas transport initiation requires $u_{\ast}$ to exceed the upper limit of this intermediate range.

\section*{Discussion}

This work presents the first systematic calculation of the turbulent wind flow over a slice of a transverse dune for different values of the wind shear velocity, $u_{\ast}$. Using computational fluid dynamics, we have shown, for the first time, that the length of the separation streamline at the lee of the dune increases with $u_{\ast}$, and that the separation streamline has an angle with the ground at the reattachment point which decreases with $u_{\ast}$. These results will have implications for morphodynamic modeling of sand dunes, since the shape of the separating streamline affects the analytical solution for the shear stress over the windward side of the dune \citep{Kroy_et_al_2002}.

Moreover, our calculations allowed us to estimate, for the first time, the values of upwind shear velocity required to induce transport of sand in the direction opposite to dune migration within the separation bubble. It is important to remark that our estimation of the minimal upwind shear velocity required to induce sustained saltation within the bubble (that is, our estimation of $u_{{\ast}{\mathrm{b}}}$) did not account for the transient distance required for the saltation flux to adapt to a change in flow conditions. This transient distance, called {\em{saturation length}} ($L_{\mathrm{sat}}$), is of the order of 50 cm and depends only marginally on the wind velocity \citep{Andreotti_et_al_2010}. Therefore, saltation transport in the reverse direction within the separation bubble can be sustained only if the magnitude of ${\tau}_x$ is above the impact threshold shear stress, ${\tau}_{\mathrm{t}} \approx {\rho}_{\mathrm{fluid}} u_{\mathrm{t}}^2$, over a distance of about 50 cm. Indeed, we find that, for an upwind shear velocity $u_{\ast} = 0.4\,$m$/$s, the value of ${\tau}_x$ varies by only $0.8\%$ over a distance of 50 cm upwind from the position $l_{\mathrm{rev}}$. In particular for this value of $u_{\ast}$, which is close to our estimate of $u_{{\ast}{\mathrm{b}}}$, $|{\tau}_x|$ is above ${\tau}_{\mathrm{t}}$ over an upwind distance of about 3\,m from the position $l_{\mathrm{rev}}$, that is almost six times $L_{\mathrm{sat}}$. Therefore, we conclude that including the effect of the saturation length affects only marginally our estimation of the lower bound for the green area in Fig.~\ref{fig:ustar_sepbub}.

By using Eq.~(\ref{eq:wind_profile}), we obtain that the wind velocity required for inducing sustained (reverse) transport in the separation bubble is about $5.0\,$m$/$s (at a height of 1~m), while an upwind flow velocity of about $6.3\,$m$/$s is required for initiating reverse transport in the dune wake. Our results shed light on the occurrence of sand transport gusts in the direction opposite to dune migration trend within the separation bubble, as reported from field observations \citep{Walker_and_Nickling_2002,Walker_and_Shugar_2013}. In fact, the wind velocity can vary considerably over time, as illustrated by the measurements of Ref.~\cite{Duran_et_al_2011}. These authors reported measurements of the surface wind speed within a barchan dune field over about 2 hours using an anemometer which resolved changes in wind speed within intervals of 1~s. The authors found strong, rapid fluctuations in wind velocity within the broad range between 4~m$/$s and 11~m$/$s \citep{Duran_et_al_2011}. Such time fluctuations of the upwind shear velocity combined with the so-called hysteresis effect in sand transport \citep{Kok_2010} --- which means that although transport initiates only if the shear velocity exceeds $u_{{\mathrm{ft}}}$ it can persist even if the shear velocity decreases down to a smaller value $u_{\mathrm{t}}$ --- provide an explanation for the sporadic gusts of sand transport within the separation bubble. 

We note that the relevance of wind fluctuations for the occurrence of transport in the separation bubble should depend on the shape of the dune and also on the values of upwind shear velocity, which can further change over different dune fields. Moreover, the geomorphological implications of the reverse transport within the separation bubble are still uncertain \citep{Walker_and_Nickling_2002,Walker_and_Shugar_2013} and should be investigated with modeling. Indeed, in real conditions the wind is not only varying in speed but also in direction \citep{Walker_and_Shugar_2013}. Thus, future work is required for improving our understanding of the secondary flow patterns on the wake of three-dimensional dune shapes, including the effect of wind directions oblique to the symmetry axis of the dune.

It would be interesting to perform the calculations of the present work considering three-dimensional dune shapes and to investigate the role of $u_{\ast}$ for the patterns of secondary flow in the lee of barchans and transverse dunes. Moreover, the calculations should be repeated for Martian conditions, in particular for investigating the effect of $u_{\ast}$ on the reversing flow within the separation bubble of Martian dunes. Indeed, the value of $u_{\mathrm{ft}}/u_{\mathrm{t}}$ on Mars can be much larger than it is on Earth, and can reach values close to ten \citep{Almeida_et_al_2008,Kok_2010}. Future modeling should account for results reported in this paper in order to investigate their implications for the dune morphodynamics.

\begin{figure}[htpb]
\begin{center}
\includegraphics[width=0.8 \columnwidth]{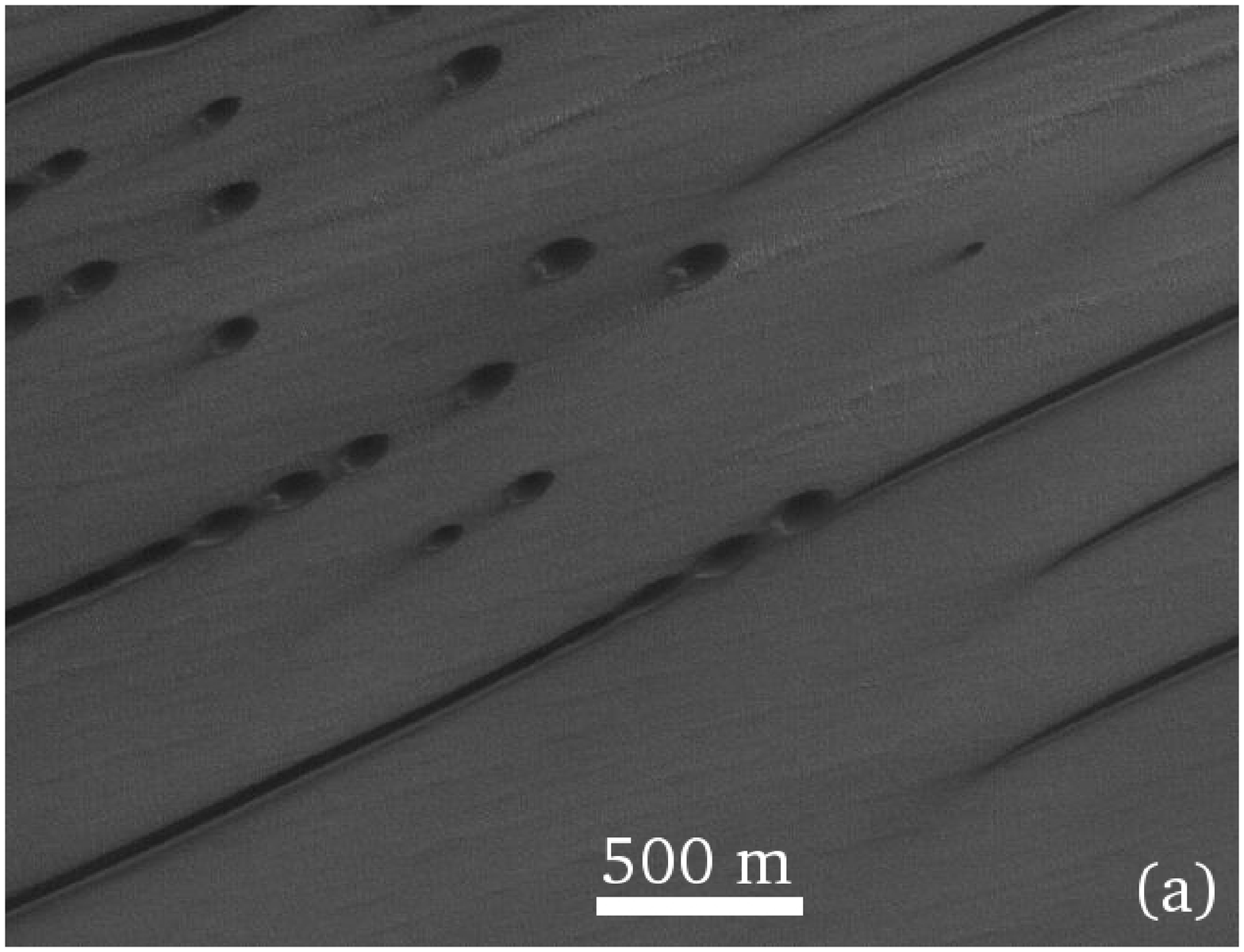}
\includegraphics[width=0.8 \columnwidth]{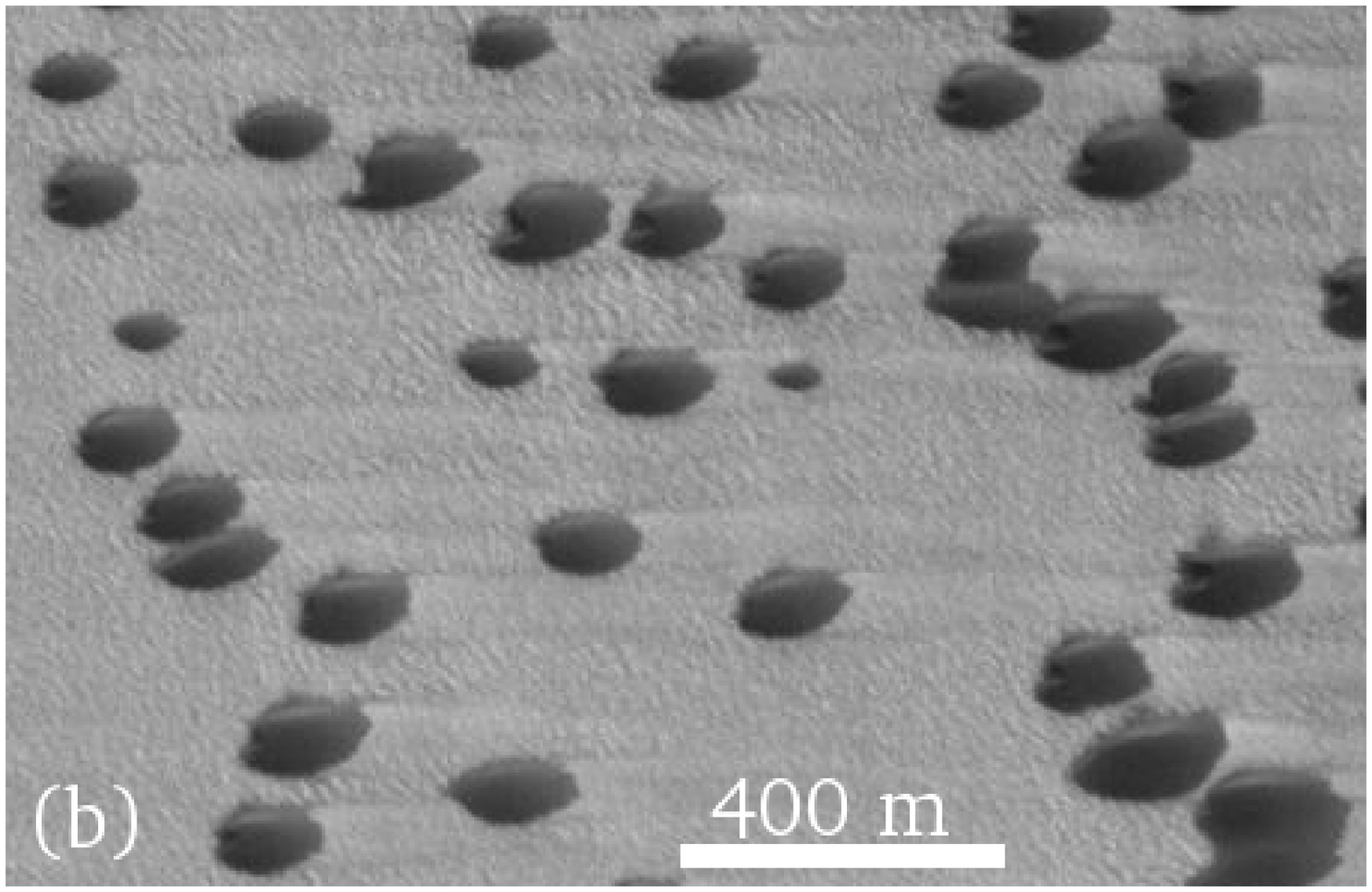}
\includegraphics[width=1.0 \columnwidth]{fig8c_d.eps}
\caption{{\bf{Sand induration can explain the origin of exotic dome-like dunes on Mars.}} Dunes at Chasma Boreale dune field near (a) $84.2^{\circ}$N, $37.9^{\circ}$W and (b) $84.53^{\circ}$N, $0.09^{\circ}$W. Images credit: NASA/JPL/MSSS. In (c) we see the transverse dune profile with the height $h$ rescaled by $H$ (the height at the brink), whereas the downwind position $x$ is rescaled by the dune's along-wind width $L$. In (d) we see the surface which results after the sand incoming from the upwind is deposited within the separation bubble calculated with $u_{\ast} = 0.9\,$m$/$s. By assuming that this surface is indurated, a rounded dune shape is obtained, which does not elongate further to form a straight dune.}
\label{fig:Chasma_Boreale}
\end{center}
\end{figure}

In particular, understanding some of the exotic dune types occuring in extraterrestrial environments requires an accurate description of the shape of the separation bubble and the wind profile within the zone of recirculating flow \citep{Kok_et_al_2012}. In the north polar region of Mars known as Chasma Boreale, unusual dune forms occur which do not find terrestrial counterparts. The unusual dune morphologies are straight linear dunes and rounded barchans or ``dome-like'' dunes with incipient slip-face, both dune shapes occurring side-by-side (cf.~Fig.~\ref{fig:Chasma_Boreale}a). 
From the orientation of the barchan (or dome-like) dunes in Chasma Boreale, it is clear that the straight dunes are {\em{longitudinal}} dunes, that is, they elongate along the wind direction. However, on Earth, longitudinal dunes form in areas of bimodal wind regimes and usually display a characteristic meandering \citep{Pye_and_Tsoar_1990,Tsoar_2001}. Moreover, it was shown that a longitudinal dune subjected to an unimodal wind is unstable and decays into a longitudinal chain of barchans \citep{Reffet_et_al_2010,Parteli_et_al_2011}. Thus, if the wind regime at Chasma Boreale is unimodal (which is the wind regime consistent with the occurrence of the barchans), then the linear dunes shouldn't occur. 

Recently it was suggested that a different mechanism might be at play in the formation of the dunes at Chasma Boreale, namely sand induration or cementation by salts or moisture in the dune \citep{Schatz_et_al_2006}. According to this hypothesis, the straight linear dunes at Chasma Boreale would form from deposition of sand in the wake of the (indurated) barchan dunes. More precisely, according to this picture, sediment transported from the upwind over the windward side of the (non-erodible) barchan is trapped within the separation bubble. As the deposited sediment undergoes cementation, the dune elongates. However, morphodynamic simulations \citep{Schatz_et_al_2006} produced indurated dome-like dunes which did not elongate further to form straight linear dunes. 

While the simulations of Ref.~\cite{Schatz_et_al_2006} were performed with a constant wind speed, here we tested the hypothesis that an indurated dome-like dune could elongate into a straight longitudinal dunes if the wind strength (and thus the separation bubble length) oscillates in time. Specifically, the volume of sand accumulated in the separation bubble during the period of strong wind could serve as source of sand for downwind transport and elongation of the dune in the direction parallel to the wind during the period of low wind strength (and thus short separation bubble length).

However, we could not produce elongated dunes using two-dimensional simulations of the dune's longitudinal profile with oscillating wind strength. The initial profile in our simulation is the transverse dune of Fig.~\ref{fig:streamlines}, which we use to compute the separation streamline for a large value of $u_{\star}$. Next, we take the envelope comprising the dune profile and the separation bubble as a new (indurated) sand surface, for which we then compute the average turbulent wind flow using a much lower value of $u_{\star}$. However, since this new dune profile is smooth (cf.~Fig.~\ref{fig:streamlines}), flow separation does not set in anymore, regardless of further oscillations in the wind speed. The predictions of our simulation is that the final dune shape is the rounded (dome-like) dune (Fig.~\ref{fig:Chasma_Boreale}b), rather than the straight dune.

Indeed, it was also conjectured in Ref.~\cite{Tsoar_2001} that three-dimensional flow patterns --- due to deflection of the flow on both flanks of the dune into the separation bubble --- at the dune lee contribute an essential factor for driving the elongation of the dune into a straight longitudinal dune. The result of our calculations apparently corroborate the hypothesis of Ref.~\cite{Tsoar_2001} that the elongation of the Chasma Boreale straight dunes cannot be understood from two-dimensional calculations (in contrast to the formation of transverse dunes). Therefore, future modeling of the Chasma Boreale straight linear dunes should be conducted using calculations of the turbulent wind flow over the three-dimensional dune shape following the present work.

\section*{Methods}\label{sec:model}

In our simulations, we consider that the fluid (air) is incompressible and Newtonian, which is a reasonable assumption since the flow velocities involved in our simulations are much smaller than the speed of sound, whereas the average turbulent wind field over the terrain is computed as described in previous works \citep{Herrmann_et_al_2005,Schatz_and_Herrmann_2006}. We adopt the FLUENT Inc. commercial package (version 6.1.25) in order to solve the Reynolds-averaged Navier-Stokes equations with the standard $\kappa-\epsilon$ model, which is used to describe turbulence.

\subsection*{\label{sec:boundary_conditions}Boundary conditions}

The logarithmic wind profile given by Eq.~(\ref{eq:wind_profile}) is imposed as boundary condition at the inlet of the channel, and the shear velocity $u_{\ast}$ is the only parameter we change in the calculations. At the outlet of the channel, a constant pressure ($P=0$) is imposed in order to generate a pressure gradient in the flow direction. The no-slip boundary condition is applied to the entire solid-fluid interface comprising the dune surface and the bottom wall, whereas for the top wall we set both components of the shear stress equal to zero, which means that the top wall moves downwind with velocity equal to the flow velocity at the height of the wall (see also Refs.~\cite{Herrmann_et_al_2005,Almeida_et_al_2006,Almeida_et_al_2008}).

\subsection*{\label{sec:discretisationscheme}Discretisation scheme and specification of the turbulence model}

We solve the time-averaged (or Reynolds-averaged) Navier-Stokes equations for the wind flow over the dune in the fully-developed turbulence regime. We use the so-called $k-\epsilon$ model with renormalization group (RNG) extensions, which is known to yield most accurate results for flow separation \citep{Lien_and_Leschziner_1994,Bradbrook_et_al_1998,Walker_and_Nickling_2002,Schatz_and_Herrmann_2006}. We choose the default pressure-velocity coupling scheme (``SIMPLE'') of the FLUENT software and use its preselected values of parameters. We also choose the software's default option ``standard wall functions''. These wall functions apply the wall boundary conditions to all solution variables of the $k-\epsilon$ model that are consistent with the logarithmic law for the wind velocity (Eq.~(\ref{eq:wind_profile})) along the entire bottom wall of the channel \citep{Launder_and_Spalding_1974}. We employ the second-order upwind scheme for the turbulence kinetic energy, turbulence dissipation rate and velocity, while for the pressure the second-order discretisation scheme is adopted. We use a triangular grid with average spacing $5\,\text{cm}$, which is refined near the dune-fluid interface and in the wake region behind the dune close to the ground. 

The initial condition is such that pressure and velocity are set to zero for all values of $x$ and $z$, except at the left wall ($x=0$), at which the logarithmic profile for the wind velocity, i.e. Eq.~(\ref{eq:wind_profile}), is imposed. The transport equations for the RNG $k-\epsilon$ model \citep{FLUENT_user_guide} are then numerically solved iteratively until the convergence criteria are fulfilled. In our simulations, the convergence criteria are defined in terms of residuals, which give a measure of the degree to which the conservation equations are satisfied throughout the flow field. We consider that convergence is achieved when the normalized residuals for both velocity components fall below $10^{-7}$ and when the normalized residuals for both $k$ and $\epsilon$ fall below $10^{-4}$. \\

\noindent
{\large{\bf{Acknowledgments}}} \newline
We thank Haim Tsoar for the image of his experiment with smoke over a barchan dune (Fig.~\ref{fig:images}c), as well as for stimulating discussions about straight linear dunes at Chasma Boreale, which motivated us to perform this study. We also thank Ingo Rehberg for providing us with the image of the subaqueous transverse dune in Fig.~\ref{fig:images}d. We acknowledge discussions with Giles Wiggs, Murilo Almeida, Antoine Wilbers, Matthew Baddock, Thomas P\"ahtz, Kai Huang and Christof Kr\"ulle. This work was supported in part by CAPES, CNPq and FUNCAP (Brazilian agencies), by the German Research Foundation (DFG) through the Cluster of Excellence ``Engineering of Advanced Materials'' and by Swiss National Foundation Grant NF 20021-116050/1 and ETH Grant ETH-10 09-2. We acknowledge support by Deutsche Forschungsgemeinschaft and Friedrich-Alexander-Universit\"at Erlangen-N\"urnberg within the funding programme Open Access Publishing. \\

\noindent
{\large{\bf{Author contributions}}} \newline
A.D.A., E.J.R.P., T.P., J.S.A.Jr. and H.J.H. contributed equally to the paper. \\

\noindent
{\large{\bf{Additional information}}} \newline
Competing financial interests: The authors declare no competing financial interests.

\end{document}